\documentclass[sigconf]{acmart}

\AtBeginDocument{%
  \providecommand\BibTeX{{%
    \normalfont B\kern-0.5em{\scshape i\kern-0.25em b}\kern-0.8em\TeX}}}
    
\copyrightyear{2021}
\acmYear{2021}
\setcopyright{acmcopyright}\acmConference[CHI '21]{CHI Conference on Human Factors in Computing Systems}{May 8--13, 2021}{Yokohama, Japan}
\acmBooktitle{CHI Conference on Human Factors in Computing Systems (CHI '21), May 8--13, 2021, Yokohama, Japan}
\acmPrice{15.00}
\acmDOI{10.1145/3411764.3445260}
\acmISBN{978-1-4503-8096-6/21/05}

\usepackage{tabulary}
\usepackage{subcaption}
\usepackage{graphicx}
\usepackage{csquotes}
\usepackage{multirow}

\begin{document}

\title{Human Perceptions on Moral Responsibility of AI: A Case Study in AI-Assisted Bail Decision-Making}

\author{Gabriel Lima}
\email{gabriel.lima@kaist.ac.kr}
\affiliation{%
  \institution{School of Computing, KAIST}
}
\affiliation{%
  \institution{Data Science Group, IBS}
  \country{Republic of Korea}
}

\author{Nina Grgić-Hlača}
\email{nghlaca@mpi-sws.org}
\affiliation{%
  \institution{Max Planck Institute for Software Systems}
}
\affiliation{%
  \institution{Max Planck Institute for Research on Collective Goods}
  \country{Germany}
}

\author{Meeyoung Cha}
\email{mcha@ibs.re.kr}
\affiliation{%
  \institution{Data Science Group, IBS}
}
\affiliation{%
  \institution{School of Computing, KAIST}
  \country{Republic of Korea}
}








\renewcommand{\shortauthors}{Lima et al.}

\begin{abstract}

How to attribute responsibility for autonomous artificial intelligence (AI) systems' actions has been widely debated across the humanities and social science disciplines. This work presents two experiments ($N$=200 each) that measure people's perceptions of eight different notions of moral responsibility concerning AI and human agents in the context of bail decision-making. Using real-life adapted vignettes, our experiments show that AI agents are held causally responsible and blamed similarly to human agents for an identical task. However, there was a meaningful difference in how people perceived these agents' moral responsibility; human agents were ascribed to a higher degree of present-looking and forward-looking notions of responsibility than AI agents. We also found that people expect both AI and human decision-makers and advisors to justify their decisions regardless of their nature. We discuss policy and HCI implications of these findings, such as the need for explainable AI in high-stakes scenarios.

\end{abstract}



\maketitle

\section{Introduction}
Who should be held responsible for the harm caused by artificial intelligence (AI)? This question has been debated for over a decade since Matthias' landmark essay on the \textit{responsibility gap} of autonomous machines~\cite{matthias2004responsibility}. This gap is posed by highly autonomous and self-learning AI systems. Until now, scholars in multiple disciplines, including ethics, philosophy, computer science, and law, have suggested possible solutions to this moral and legal dilemma. Optimistic views proclaim that the gap can be bridged by proactive attitudes of AI designers, who should readily take responsibility for any harm~\cite{nyholm2018attributing,champagne2015bridging}. Some even propose to hold AI systems responsible per se~\cite{stahl2006responsible}, viewing human-AI collaborations as extended agencies~\cite{gunkel2017mind,hanson2009beyond}. In contrast, pessimistic views question whether this gap can be bridged at all, since there might not exist appropriate subjects of retributive blame~\cite{danaher2016robots} nor it makes sense to hold inanimate and non-conscious entities responsible for their actions~\cite{sparrow2007killer,torrance2008ethics,bryson2017and}. 

Most research on the responsibility gap has been normative in that they prescribed ethical principles and proposed solutions. However, there is a growing need for practical and proactive guidelines; as Mittelstadt puts it, ``principles alone cannot guarantee ethical AI''~\cite{mittelstadt2019principles}. Some even argue that normative approaches are inappropriate as they can hurt AI's adoption in the long run~\cite{bonnefon2020moral}. In contrast, relatively little attention has been paid to understanding the public's views on this issue, who are likely the most affected stakeholder when AI systems are deployed~\cite{rahwan2018society}.

We conducted two survey studies ($N$=200 each) that collect the public perception on moral responsibility of AI and human agents in high-stakes scenarios. We approached the pluralistic view of responsibility and considered eight distinct notions compiled from philosophy and psychology literature. Real-life adapted vignettes of AI-assisted bail decisions were used to observe how people attributed specific meanings of responsibility to \textit{i}) AI advisors vs. human advisors and \textit{ii}) AI decision-makers vs. human decision-makers. Our study employed a within-subjects design where all participants were exposed to a diverse set of vignettes addressing distinct possible outcomes from bail decisions.  

Our findings suggest that the eight notions of responsibility considered can be re-grouped into two clusters: one encompasses \emph{present-looking and forward-looking} notions (e.g., responsibility-as-task, as-power, as-authority, as-obligation), and the other includes \emph{backward-looking} notions (e.g., blame, praise, liability) and \emph{causal determinations}. We discuss how theories of moral responsibility can explain these clusters.

In comparing AI agents against human agents, we found a striking difference in the way people attribute responsibility. A substantially higher degree of the present- and forward-looking notions were attributed to human agents than AI agents. This means that AI agents were assigned the responsibility to complete and oversee the same task to a lesser extent than human agents. No difference, however, was observed for the backward-looking responsibility notions. This finding suggests that blame, liability, and causal responsibility were ascribed equally to AI and human agents, despite electronic agents not being appropriate subjects of liability and blame~\cite{sparrow2007killer,danaher2016robots,bryson2017and}. In addition to these findings, we found that people expect both human and AI agents to justify their decisions.

The findings of this study have several implications for the development and regulation of AI. Using the proposition of morality as a human-made social construct that aims to fulfill specific goals~\cite{theodorouartificial,stahl2006responsible}, we highlight the importance of users and designers taking responsibility for their systems while being held responsible for any norm-violating outcomes. We also discuss the possibility of holding AI systems responsible per se~\cite{lima2020responsible} \emph{alongside} other human agents, as a possible approach congruent to the public opinion.

\section{Background}
\subsection{Theories of (Moral) Responsibility}

Theories of moral responsibility date back to Aristotle, who argued that an entity should satisfy both freedom and epistemic conditions to appropriately be ascribed to moral responsibility. Agents must act freely, without coercion, and understand their actions. Although recent scholarly work does not directly challenge these Aristotelian conditions, they argue that moral responsibility cannot be explained as a single concept, but that it involves a relatively pluralistic definition of what it means to hold someone morally responsible~\cite{van2015moral,shoemaker2011attributability}.

Scanlon~\cite{scanlon2000we} has proposed moral responsibility to be a bipartite concept. One is that there is an account of \emph{being} responsible in rendering an agent worthy of moral appraisal. Another is that it is also possible to \emph{hold} one responsible for specific actions and consequences. Expanding this bipartite concept, Shoemaker~\cite{shoemaker2011attributability} has proposed three different concepts of moral responsibility: attributability, answerability, and accountability. Various other definitions have been proposed~\cite{van2015moral}, including structured notions of what responsibility is~\cite{vincent2011structured} and how they are connected~\cite{bovens1998quest,duff2007answering}.

Attributing responsibility to an entity can be both descriptive (e.g., causal responsibility) and normative (e.g., blameworthiness). For the former, one might ask if an agent \emph{is} responsible for an action or state-of-affairs, while the latter concerns whether one \emph{should} attribute responsibility to an agent. Responsibility can also be divided into backward-looking notions if they evaluate a past action and possibly lead to reactive attitudes~\cite{wallace1994responsibility}, or forward-looking notions if they prescribe obligations.

Responsibility can take many forms. It not only addresses the moral dimension of society but also tackles legal concepts and other descriptive notions. One can be held legally responsible (i.e., liable) regardless of their moral responsibility, as in the case of strict or vicarious liability. Stating that an agent is causally responsible for a state-of-affairs does not necessarily prescribe a moral evaluation of the action. 

Holding an agent ``responsible'' fulfills a wide range of social and legal functions. Legal scholars state that punishment (which could be seen as a form of holding an agent responsible, e.g., under criminal liability) aims to reform the wrongdoers, deter re-offenses and similar actions, and resolve retributive sentiments~\cite{asaro201111,twardawski2020all}. Previous work has addressed how and why people assign responsibility to various agents. The general public might choose to hold a wrongdoer responsible for restoring moral coherence~\cite{clark2015moral} or reaffirming a communal moral values~\cite{wenzel2006we}. Psychological research indicates that people base much of their responsibility attribution on retributive sentiments rather than deterrence~\cite{carlsmith2002we}, while overestimating utilitarian goals in their ascription of punishment (i.e., responsibility)~\cite{carlsmith2008justifying}. Intentionality also determines how much responsibility is assigned to an entity~\cite{monroe2017two}; people look for an intentional agent to hold responsible and infer other entities' intentionality upon failure to find one~\cite{gray2014myth}. 

\subsection{Techno-Responsibility Gaps}

AI systems and robots are being widely adopted across society. Algorithms are used to choose which candidate is most fit for a job position~\cite{zhu2018person}, decide which defendants are granted bail~\cite{dressel2018accuracy}, guide health-related decision~\cite{obermeyer2019dissecting}, and assess credit risk~\cite{huang2004credit}. AI systems are often embedded into robots or machines, such as autonomous vehicles~\cite{bonnefon2016social} and robot soldiers~\cite{asaro2006should}. A natural question here is: if an AI system or a robot causes harm, who should be held responsible for their actions and consequences?

In answering this question, some scholars have defended the existence of a (techno-)responsibility gap~\cite{matthias2004responsibility} for autonomous and self-learning systems.\footnote{Scholars also raise doubt on the existence of techno-responsibility gaps, arguing that moral institutions are dynamic and flexible and can deal with these new technological artifacts~\cite {tigard2020there,kohler2017technology}.} The autonomous component of AI and robots challenges the control condition of responsibility attribution. Simultaneously, their self-learning capabilities and opacity do not allow users, designers, and manufacturers to foresee consequences. Similarly to the ``problem of many hands'' in the assignment of responsibility to collective agents~\cite{van2015moral}, AI and robots suffer from the ``problem of many things,'' i.e., current systems are composed of various interacting entities and technologies, making the search for a responsible entity harder~\cite{coeckelbergh2019artificial}. Scholars have extensively discussed the assignment of responsibility for autonomous machines' actions and have expanded this gap to more specific notions of responsibility~\cite{asaro2016liability,koops2010bridging,beck2016problem} and its functions~\cite{danaher2016robots,limachi2020}.

Although a clear separation is fuzzy, one may find two schools of thought on the responsibility gap issue. One side argues that designers and manufacturers should take responsibility for any harm caused by their ``tools.''~\cite{dignum2017responsible,bryson2017and} Supervisors and users of these systems should also take responsibility for their deployment, particularly in consequential environments like the military as argued by Champagne and Tonkens~\cite{champagne2015bridging}. The exercise of agency by these systems can be viewed as a human-robot collaboration, in which humans supervise and manage the agency of AI and robots~\cite{nyholm2018attributing}. Humans should focus on their relationship to the patients of their responsibility to answer for the actions of autonomous systems~\cite{coeckelbergh2019artificial}. Likewise, other authors argue that society should hold humans responsible because doing so for a machine would be meaningless as it does not understand the consequences of their actions or the reactive attitudes towards them~\cite{sparrow2007killer,torrance2008ethics}, possibly undermining the definition of responsibility~\cite{hakli2016robots}.

On the opposite side, some scholars propose autonomous systems could be held responsible per se~\cite{lima2020responsible}. From a legal perspective, non-human entities (e.g., corporations) can be held responsible for any damage that they may cause~\cite{van2018we}. These scholars often view these human-AI collaborations as extended agencies where all entities should be held jointly responsible~\cite{hanson2009beyond,gunkel2017mind}. AI and robots are part of the socio-technological ensemble, in which responsibility can be distributed across multiple entities with varying degrees~\cite{dodig2008sharing}. These proposals arguably contribute to legal coherence~\cite{turner2018robot}, although it could also lead to various repercussions in moral and legal institutions~\cite{beck2016problem}. 
Empirical findings indicate that people attribute responsibility to these systems~\cite{awad2020drivers,limachi2020}, although to a lesser extent than human agents. According to some scholars, holding AI and robots responsible per se could fulfill specific social goals~\cite{coeckelbergh2009virtual} and promote critical social functions~\cite{bjornsson2012explanatory,stahl2006responsible}.

The regulation of AI and robots poses new challenges to policymaking, as in the previously introduced techno-responsibility gap, which society must discuss at large~\cite{coeckelbergh2019artificial}. The ``algorithmic social contract'' requires inputs from various stakeholders, whose opinion should be weighed for the holistic crafting of regulations~\cite{rahwan2018society}. It is crucial to understand how people perceive these systems before their wide deployment~\cite{richards2016should}. Our responsibility practices depend on folk-psychology~\cite{brozek2017legal} (i.e., how people perceive the agents involved in social practices~\cite{stahl2006responsible}). Literature exists on the public perception of moral and legal issues concerning AI~\cite{awad2020drivers,limachi2020,awad2018moral}. However, little data-driven research has collected public opinion on how responsibility should be attributed for AI and robots' actions.

\subsection{Responsibility, Fairness, Trust in HCI Literature}
\label{sec:hci}

A growing number of HCI research has been devoted to understanding how people perceive algorithmic decisions and their consequences in society. For instance, Lee et al. studied people's perception of trust, fairness, and justice in the context of algorithmic decision-making~\cite{lee2018understanding,lee2019procedural} and proposed how to embed these views into a policymaking framework~\cite{lee2019webuildai}. 
Other scholars explored people's perceptions of procedural~\cite{grgic2018human} and distributive~\cite{saxena2018fairness, srivastava2019mathematical} aspects of algorithmic fairness and studied how they relate to individual differences~\cite{wang2020fairness, pierson2017gender, grgic2020dimensions}.
Nonetheless, little attention is paid to the public attribution of (moral) responsibility to stakeholders (e.g., ~\cite{lee2018understanding,grgic2019human,robert2020designing}), particularly the prospect of responsibility ascription to the AI system per se. The current study contributes by addressing the public perception of algorithmic decision-making through the lens of moral responsibility.

Existing studies addressing how users might attribute blame to automated agents have mostly focused on robots. For instance, Malle et al. observed that people's moral judgments between human and robotic agents differed in that respondents blamed robots to a more considerable extent had they not taken a utilitarian action~\cite{malle2015moral}. 
Furlough et al. found that respondents attributed similar levels of blame to robotic agents and humans when robots were described as autonomous and at the same time the leading cause of harm~\cite{furlough2019attributing}.
However, these studies and many others~\cite{li2016trolley,wachter2018explorative,kim2006should} tackle a singular notion of responsibility related to blameworthiness~\cite{van2015moral}. The present research explores multiple notions of moral responsibility of both human and AI agents involved in decision-making.


%


\section{Methodology}\label{sec:methods}
\subsection{Algorithmic Decision-Making}

AI-based algorithms are now used to assist humans in various scenarios, including high-stakes tasks such as medical diagnostics~\cite{esteva2017dermatologist} and bail decisions~\cite{propublica_story}. These algorithms do not make decisions themselves, but rather ``advise'' humans in their decision-making processes. One such algorithm is the COMPAS (Correctional Offender Management Profiling for Alternative Sanctions) tool, used by the judicial system in the US to assist bail decisions and sentencing~\cite{propublica_story}. Several studies have analyzed the fairness and bias aspects of this risk assessment algorithm, e.g.,~\cite{dressel2018accuracy, berk2018fairness, grgic2019human}.

This study makes use of publicly available COMPAS data released by ProPublica~\cite{propublica_story} and considers the machine judgments as either an AI advisor (later in Study 1) or an AI decision-maker (in Study 2). As stimulus material, we use real-world data obtained from a previous analysis of the tool~\cite{propublica_story}, which focused on its application in bail decision-making. This dataset contains information about 7,214 defendants subjected to COMPAS screening in Broward County, Florida, between 2013 and 2014.

We use 100 randomly selected cases from this dataset, the corresponding bail suggestions, and information about whether the defendant re-offended within two years of sentencing. The sampled data was balanced concerning these variables. Each defendant's COMPAS score ranges from 1 to 10, with ten indicating the highest risk of re-offense or nonappearance in court. In this study, scores 1 to 5 were labeled  ``grant bail'' and 6 to 10 were labeled ``deny bail.''

\subsection{The Plurality of Responsibility} 
\label{sec:plurality_of_responsibility}


Ascribing responsibility is a complex moral and legal practice that encompasses various functions, entities, and social practices~\cite{mulligan2017revenge,stahl2006responsible}. Responsibility has multiple distinct meanings depending on its purpose and requirements. 
%
%
The current study revisits eight notions of responsibility compiled from psychology and philosophy. All of these notions originated from Van de Poel's work~\cite{van2011relation,van2015moral}, except for responsibility-as-authority and as-power, which comes from Davis's discussion on professional responsibility~\cite{davis2012ain}. We complement these notions with a wide range of literature ranging from philosophical theories of moral responsibility (e.g.,~\cite{shoemaker2011attributability,scanlon2008moral}) to approaches in the context of AI systems (e.g.,~\cite{coeckelbergh2019artificial,torrance2008ethics}). Although not exhaustive (e.g., we have not addressed virtue-based notions of responsibility as they cannot be easily adapted to AI systems), we highlight how our work differs from previous HCI approaches.

\begin{itemize}
    \item \textbf{Responsibility-as-{obligation}:} \\
    \textit{E.g., ``The (agent) should ensure that the rights of the defendant are protected.''} \\ 
    One could be held responsible-as-obligation through consequentialist, deontological, and virtue-based routes~\cite{van2015moral}. While an entity could be attributed such meaning of responsibility based on pre-determined consequentialist distribution principles, the latter two routes presuppose the agent's initiative or promise to see to it that a specific state-of-affairs is brought about. This notion differs from responsibility-as-task as it does not imply that one should be the agent to bring about a specific state-of-affairs, but rather indicates that one \emph{should} fulfill its supervisory duties in the process.
    
    \vspace*{1mm}
    \item \textbf{Responsibility-as-{task}:} \\
    \textit{
    ``It is the (agent)'s task to protect the defendant's rights.''} \\
    This descriptive notion of responsibility ascribes a specific task to an entity. These assignments do not necessarily define a moral obligation per se~\cite{van2011relation} and are often accompanied by the understanding that an entity \emph{has} to do something by itself~\cite{davis2012ain}. In our experimental design, we highlight the agent's acting role in completing its task.

    \vspace*{1mm}
    \item \textbf{Responsibility-as-{authority}:} \\
    \textit{
    ``The (agent) has the authority to prevent further offenses.''} \\
    To be responsible-as-authority implies that one \emph{is} in charge of a specific action or state-of-affairs. This notion has also been posed as "responsibility-as-office" by Davis~\cite{davis2012ain} in the context of engineers' professional responsibility. An important aspect of responsibility-as-authority is the possibility of delegating other complementing notions, such as responsibility-as-task, to other agents. We address this meaning of responsibility by explicitly indicating that the agent has the authority in bail decisions.

    \vspace*{1mm}
    \item \textbf{Responsibility-as-{power}:} \\
    \textit{
    ``The (agent) has the skills needed to protect the rights of the defendant.''} \\
    If an entity \emph{has} the skills needed to bring about an action or state-of-affairs, one might ascribe it responsibility-as-power~\cite{davis2012ain}. In other words, having the ability, in terms of competency, knowledge, or expertise, might lead to the assignment of this notion of responsibility.
    
    \vspace*{1mm}
    \item \textbf{Responsibility-as-{answerability}:} \\
    \textit{
    ``The (agent) should justify their advice.''} \\
   This is related to how one's reasons for acting in a specific manner could be seen under moral scrutiny. Shoemaker proposed this notion of moral responsibility as a form of judgment of one's actions grounded in moral evaluations~\cite{shoemaker2011attributability}. Davis proposed a similar meaning of responsibility under a different name, responsibility-as-accountability~\cite{davis2012ain}, as the responsibility for explaining specific consequences. Coeckelbergh later applied this concept through a relational approach for actions and decisions made using AI~\cite{coeckelbergh2019artificial}. 

    \vspace*{1mm}
    \item \textbf{Responsibility-as-{cause}:} \\
    \textit{
    ``The (agent)'s decision led to the prevention of the re-offense.''} \\
    This meaning of responsibility has been further discussed depending on the nature of an action's consequences~\cite{davis2012ain}, e.g., being causally responsible for a positive state-of-affairs could lead to the ascription of ``good-causation.'' Causality is also an important pre-condition for other normative notions of responsibility, such as blame, as the blurring of a causal connection raises questions on the foreseeability and control of a specific action.~\cite{van2015moral,malle2014theory}

    \vspace*{1mm}
    \item \textbf{Responsibility-as-{blame/praise}:} \\
    \textit{
    ``The (agent) should be blamed for the violation of the rights of the defendant.''} 
    /
    \textit{
    ``The (agent) should be praised for the protection of the rights of the defendant.''} \\
    Blaming an entity for the consequences of their actions has been debated as adopting certain reactive attitudes towards it~\cite{wallace1994responsibility}. Scholars have also argued that to blame someone is to respond to ``the impairment of a relationship,''~\cite{chislenko2019scanlon,scanlon2008moral} especially towards its constitutive standards~\cite{shoemaker2011attributability}. Scholars have debated the possibility of ascribing blame to an automated agent and agree that doing so would not be morally appropriate~\cite{danaher2016robots,torrance2008ethics}. Regardless of this consensus, previous studies have found that people attribute a similar degree of blame to robotic and human agents under specific conditions (e.g., ~\cite{malle2015moral, furlough2019attributing}).\\
    As an opposite concept of blame, one may consider ``praise'' as a positive behavioral reinforcement~\cite{kazdin1978history} through which one conveys its values and expectations of the agent~\cite{delin1994praise}. Hence, we consider both blame and praise as responsibility notions in this research.

    \vspace*{1mm}
    \item \textbf{Responsibility-as-{liability}:} \\
    \textit{
    ``The (agent) should compensate those harmed by the re-offense.''} \\
    An entity that is ascribed this responsibility should remedy any harm caused by their actions~\cite{van2015moral}. Rather than dwelling on the discussion addressing the mental states of AI and robots and their arguable incompatibility with criminal law and its assumption of \textit{mens rea}~\cite{lima2017could,lagioia2019ai,gless2016if}, we address this notion from a civil law perspective. Scholars propose `making victims whole' as the primary goal of tort law~\cite{prosser1941handbook}, and hence, we similarly address responsibility-as-liability. 
    We also add that the idea of holding automated agents liable became prominent after the European Parliament considered adopting a specific legal status for ``sophisticated autonomous robots''~\cite{delvaux2017report}. Nevertheless, it is important to note that current AI systems cannot compensate those harmed, as they do not possess any assets to be confiscated~\cite{bryson2017and}.

\end{itemize}

\subsection{Survey Design}

\begin{figure*}
    \centering
    \includegraphics[width=\textwidth]{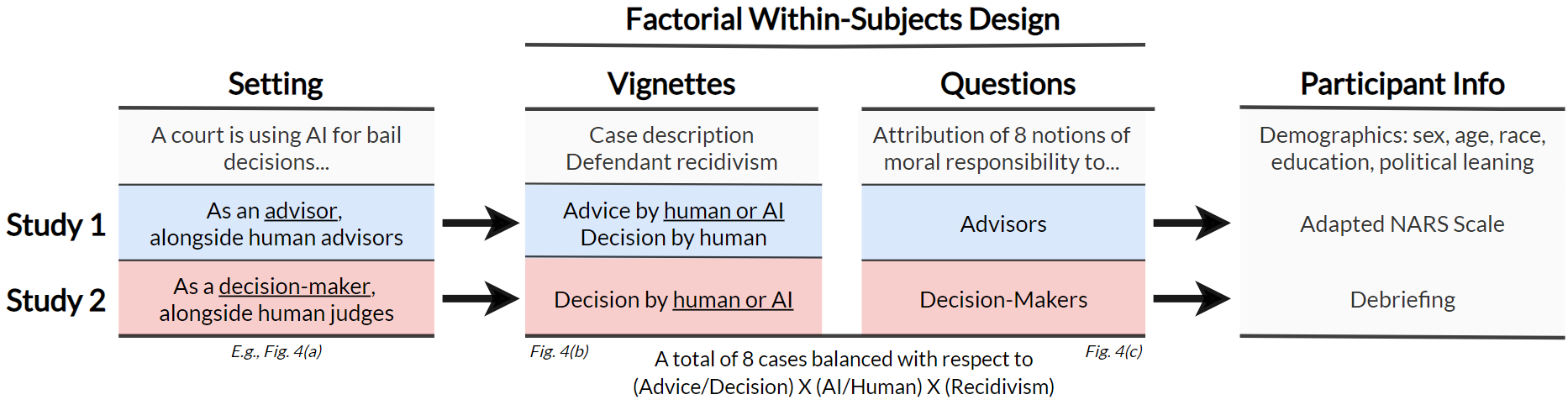}
    \caption{
    Survey Instrument.
    In Study 1, where AI advisors or human advisors assist human judges, survey respondents were asked to assign responsibility notions to AI and human advisors. In Study 2, where AI systems are decision-makers alongside human judges, survey respondents were asked to assign responsibility notions to AI and human decision-makers. Both studies employed a factorial within-subjects design that presented eight different vignettes to each respondent. Survey instruments are shown in Figure~\ref{fig:survey_instrument} of the Appendix.}
    \Description{
    Survey Instrument.
    In Study 1, where AI advisors or human advisors assist human judges, survey respondents were asked to assign responsibility notions to AI and human advisors. In Study 2, where AI systems are decision-makers alongside human judges, survey respondents were asked to assign responsibility notions to AI and human decision-makers. Both studies employed a factorial within-subjects design that presented eight different vignettes to each respondent. Survey instruments are shown in Figure~\ref{fig:survey_instrument} of the Appendix.}
    \label{fig:method}
\end{figure*}

\noindent 
$\bullet$~~\textbf{Study 1: AI as Advisor}\\
To study how the perceived responsibility for bail decisions differs when judges are \emph{advised} by the COMPAS tool or by another human judge, we considered the following scenario:

\begin{displayquote}
\emph{Imagine that you read the following story in your local newspaper:
A court in Broward County, Florida, is starting to use an artificial intelligence (AI) program to help them decide if a defendant can be released on bail before trial.
Early career judges are taking turns receiving advice from this AI program and another human judge, hired to serve as an advisor.}
\end{displayquote}

We employed a factorial survey design~\cite{wallander200925} and showed participants eight vignettes that described a defendant from the ProPublica dataset, information about who the advisor was (i.e., an AI program or a human judge), which advice they gave, what the judge's final decision was, and whether the defendant committed a new crime within the next two years (i.e., re-offended). All vignettes stated that the judges' final decision \emph{followed} the advice given, given the ProPublica dataset does not provide this information. After reading the stimulus material, respondents were asked to indicate to what extent they agreed with a set of statements, presented in random order between participants, regarding the \emph{advisor} on a 7-point Likert scale (-3 = Strongly Disagree, 3 = Strongly Agree).\footnote{
Questions related to responsibility-as-liability were shown in scenarios where \textit{i}) the defendant re-offended and the phrasing style addressed the prevention of re-offenses, 
or \textit{ii}) the defendants were denied bail and did not re-offend within two years while the statements focused on protecting their rights. The phrases tackling praise and blame were presented depending on the advice/decision and recidivism.} 
These statements aimed to capture different notions of responsibility (see Table~\ref{tab:statements} in the Appendix for the complete list). Figure~\ref{fig:method} illustrates the survey methodology. Participants were also asked two attention check questions in between vignettes.

Each participant in the study was exposed to a random subset of four cases with human advice and another four with AI advice. We ensured a balanced set was shown to each participant in terms of the advice (i.e., grant bail vs. deny bail) and recidivism. As a result, each respondent was shown one vignette of every possible combination of scenarios, encompassing eight (advice $\times$ recidivism $\times$ AI vs. human) variations. All vignettes were presented in random order to eliminate any order effect~\cite{redmiles2017summary, groves2011survey}.

Bail decisions aim to procure a balance between protecting future victims, e.g., prevent further offenses, and to impede any unnecessary burdens towards the defendant, e.g., by ensuring that their rights are protected~\cite{grgic2019human}. The latter aspect of bail decisions is related to the assumption that one is innocent until proven otherwise beyond a reasonable doubt under criminal law~\cite{dhami2015instructions}. To strike a balance between these two functions of bail decisions, we phrase statements addressing all notions of responsibility addressed in this work in two different forms: a human agent or an AI program could be held responsible for \textit{i}) (not) protecting the rights of the defendant and \textit{ii}) (not) preventing re-offense. Participants were randomly assigned to one of these treatment groups, and all statements followed the same phrasing style.

Towards the end of the survey, we asked demographic questions (presented in Table~\ref{tab:demographics}). We also gathered responses to a modified questionnaire of NARS (Negative Attitude towards Robot Scale)~\cite{syrdal2009negative}, whose subscale addressed ``artificial intelligence programs'' rather than ``robots'' to accommodate the COMPAS tool. 
\\


\noindent 
$\bullet$~~\textbf{Study 2: AI as Decision-Maker}\\
Unlike Study 1, where a human decision-maker is advised by either a human or an AI advisor, Study 2 explores a setting that has yet to be implemented in the real-world. We imagine the case where an AI algorithm \emph{makes} a bail decision by itself.
The survey instrument and experimental design are identical to Study 1, except that in the introductory text, we told participants, \emph{"The court is taking turns employing human judges and this AI program when making bailing decisions,"} and updated the phrasing of the questions to match this setting accordingly. In each vignette, participants were asked to what extent they agreed with the eight notions of responsibility regarding the \emph{decision-maker}, i.e., the AI program or the human judge, 
using the same 7-point Likert scale from Study 1. Both studies had been approved by the Institutional Review Board (IRB) at the first author's institution. \\

\noindent 
$\bullet$~~\textbf{Pilot Study for Validation: Cognitive Interview}\\
We validated our survey instruments through a series of cognitive interviews. Cognitive interviews are a standard survey methodology approach for improving the quality of questionnaires~\cite{ryan2012improving}. During the interviews, respondents accessed our web-based survey questionnaire and were interviewed by the authors while completing the survey. We utilized a verbal probing approach~\cite{willis2004cognitive}, in which we tested the respondents' interpretation of the survey questions, asked them to paraphrase the questions, and if they found the questions easy or difficult to understand and answer.

We interviewed six demographically diverse respondents. Three respondents were recruited through the online crowdsourcing platform Prolific~\cite{palan2018prolific}, while the other three were our colleagues, who had prior experience designing and conducting human-subject studies. After each interview, we iteratively refined our survey instrument based on the respondent's feedback. We stopped gathering new responses once the feedback stopped leading to new insights.
This process led to two significant changes in our survey instrument design. Firstly, we adapted the vignette presentation, which was initially adapted from previous work~\cite{dressel2018accuracy}. Our respondents unanimously stated that they found information about defendants easier to read, understand, and use when presented in a tabular format (shown in Figure~\ref{fig:survey_instrument} in the Appendix). Secondly, we rephrased some of the statements about the notions of responsibility we address in this work so that survey respondents' understanding of these concepts is similar to the definitions introduced above.

\begin{table}[t]
\centering
\small
\begin{tabulary}{\linewidth}{L|R|R|R}
\toprule
\textbf{Demographic Attribute} & \textbf{Study 1} & \textbf{Study 2} & \textbf{Census}\\
\hline
Total respondents & 203 & 197 & - \\
Passed attention checks & 200 & 194 & - \\
\hline
Women & 41.5\% & 40.7\% & 51.0\%\\
\hline
0-18 years old & - & - & 21.7\%\\
18-24 years old & 37.5\% & 30.4\% & 10.8\%\\
25-34 years old & 30.5\% & 34.0\% & 13.7\%\\
35-44 years old & 17.0\% & 18.6\% & 12.6\%\\
45-54 years old & 8.0\% & 7.2\% & 13.4\%\\
55-64 years old & 6.0\% & 6.7\% & 12.9\%\\
65+ years old & 1.0\% & 2.6\% & 14.9\%\\
Prefer not to respond & - & 0.5\% & -\\
\hline
African American & 4.5\% & 7.2\% & 13\% \\
Asian & 17.0\% & 21.6\% & 6\% \\
Caucasian & 58.5\% & 54.6\% & 61\%\\
Hispanic & 10.0\% & 8.2\% & 18\% \\
Other/Prefer not to respond & 10.0\% & 8.4\% & 4\% \\
\hline
Bachelor's Degree or above & 48.0\% & 52.0\% & 30\%\\
\hline
Liberal & 59.0\% & 58.2\% & 33\%$\dagger$\\
Conservative & 14.5\% & 18.0\% & 29\%$^\dagger$\\
Moderate & 23.5\% & 21.6\% & 34\%$^\dagger$\\
Other/Prefer not to respond & 3.0\% & 2.2\% & 4\%$^\dagger$\\
\bottomrule
\end{tabulary}
\caption{Respondents' demographics compared to the 2016 U.S. Census~\citep{census_acs} and Pew data (marked with $^\dagger$)~\citep{pew_politics}.}
\label{tab:demographics}
\end{table}

\subsection{Participants and Recruitment}

We conducted a power analysis to calculate the minimum sample size. A Wilcoxon-Mann-Whitney two-tailed test, with a 0.8 power to detect an effect size of 0.5 at the significance level of 0.05, requires 67 respondents per treatment group. Hence, we recruited 400 respondents through the Prolific crowdsourcing platform~\cite{palan2018prolific} to compensate for attention-check failures. We targeted US residents who have previously completed at least 100 tasks on Prolific, with an approval rate of 95\% or above. Each participant was randomly assigned to one of the two studies. 

The respondents' demographics are shown in Table~\ref{tab:demographics}. Prior studies of online crowdsourcing platforms have found that respondent samples tend to be younger, more educated, and consist of more women than the general US population~\cite{ipeirotis2010demographics}. Compared to the 2016 US census~\cite{census_acs}, our respondents are indeed younger and more highly educated. However, both of our studies' samples have a smaller ratio of women than the US population. Asian ethnicity is slightly over-represented in our samples. Compared to Pew Research data on the US population's political leaning~\citep{pew_politics}, our samples are substantially more liberal.

The respondents were remunerated US$\$10.5$ for taking part in the cognitive interviews and US$\$1.66$ for completing the online surveys. The cognitive interviews lasted less than 30 minutes, while the online surveys took 10.36$\pm$5.43 minutes. Hence, all study participants were paid above the US minimum wage.



\section{Results}\label{sec:results}
\subsection{Responsibility in Bail Decisions}
\label{sec:exploratory}

\begin{figure*}[t]
    \begin{subfigure}[b]{\textwidth}
         \centering
         \includegraphics[width=.95\textwidth]{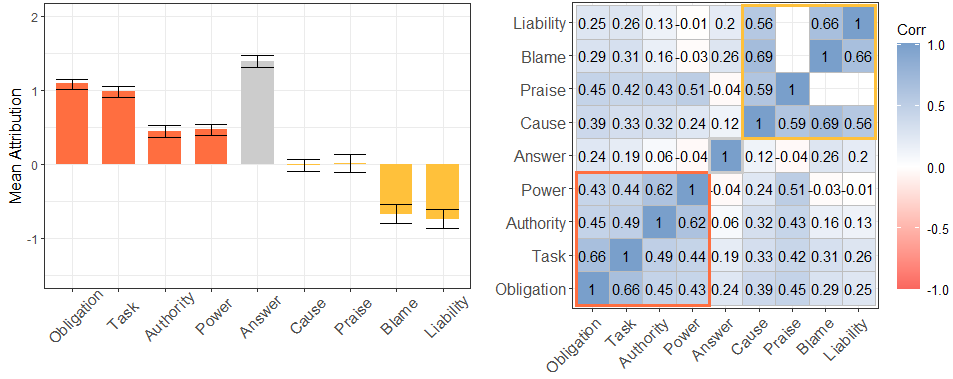}
         \caption{Study 1: Aggregate results of AI or human advisors.}
         \label{fig:exploratory1}
     \end{subfigure}
    \begin{subfigure}[b]{\textwidth}
         \centering
         \includegraphics[width=.95\textwidth]{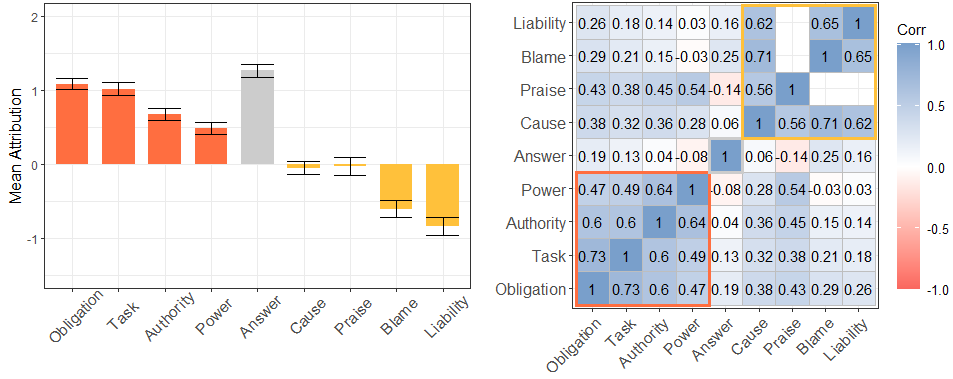}
         \caption{Study 2: Aggregate results of AI or human decision-makers.}
         \label{fig:exploratory2}
     \end{subfigure}      
     \caption{The overall attribution of responsibility to AI or human agents in bail decisions (left) and the correlation matrix across different responsibility notions (right). The $y$-axis indicates the degree to which participants attributed each notion of responsibility, based on a 7-pt Likert Scale (-3 = Strongly Disagree, 3 = Strongly Agree). 
     }
     \Description{Responsibility for bail decisions (left) and the correlation matrix between different responsibility notions (right). Participants attributed responsibility by agreeing with a set of statements proposing the attribution of each responsibility notion to AI and human agents using a 7-pt Likert Scale (-3 = Strongly Disagree, 3 = Strongly Agree). Present-looking and forward-looking notions were correlated and were attributed to a large extent. Backward-looking notions were also correlated and were attributed to a lower extent. Responsibility-as-answerability is the notion attributed the most, but it is not correlated with any other notion.}
     \label{fig:exploratory}
\end{figure*}

Figure~\ref{fig:exploratory} shows how people attributed each notion of responsibility to AI and human agents in Study 1 (on the advisor role) and Study 2 (on the decision-maker role). 

First, responsibility-as-answerability (i.e., the bar in the middle) was the notion ascribed the highest to both human and AI advisors and decision-makers, followed by responsibility-as-obligation, as-task, as-authority, and as-power (i.e., the first four bars). On the other hand, liability and blame were the least attributed responsibility notion in bail decisions. Responsibility-as-cause and praise were the most neutral notions, and their mean attribution is close to zero (i.e., the baseline) across all treatments (see Figure~\ref{fig:app} in the Appendix).

Second, Figure~\ref{fig:exploratory} shows two distinct sets of responsibility notions, where these clusters can be observed from the pairwise Spearman's correlation chart. A high correlation value indicates that those responsibility notions are perceived similarly by people. 
One group includes responsibility-as-task, as-authority, as-power, and as-obligation, all of which have positive mean values. 
The other group includes responsibility-as-cause, praise, blame, and liability. 
Responsibility-as-answerability belongs to neither of these groups.

Third, we can quantify variations across vignette conditions. Each vignette shown to participants varied in the advice given, bail decision, and recidivism, allowing us to compare across these factors. Our data show that vignettes that grant bail (as opposed to denying bail) led to a higher assignment of all responsibility notions, particularly causal responsibility and blame (see Figure~\ref{fig:app} in the Appendix). A similar effect was found depending on defendant recidivism. For instance, the first four responsibility notions were ascribed to a more considerable degree if the defendant did not re-offend. In contrast, responsibility-as-cause, blame, and liability were attributed to a lesser extent if the defendant re-offended within two years. These trends corroborate the responsibility clusters discussed above.

Finally, our study participants were also assigned to one of two different phrasing styles addressing some of the bailing decisions' objectives. Except for responsibility-as-answerability, addressing the violation or protection of a defendant's rights led to a marginally higher assignment of responsibility than the phrasing style focusing on preventing re-offenses.

\subsection{Responsibility Assignment to AI and Humans}
\label{sec:regression}

\begin{figure*}[t]
     \begin{subfigure}[b]{\textwidth}
         \centering
         \includegraphics[width=.75\textwidth]{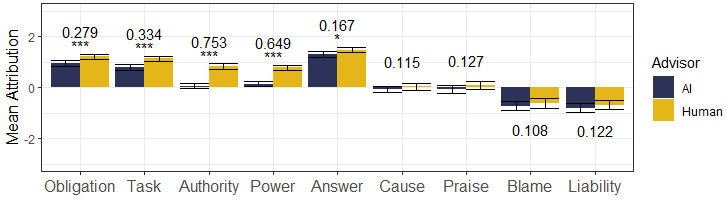}
         \caption{Study 1: AI and human advisors.}
         \label{fig:agent2}
     \end{subfigure}
    \begin{subfigure}[b]{\textwidth}
         \centering
         \includegraphics[width=.75\textwidth]{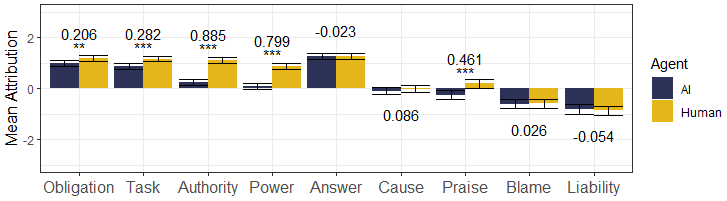}
         \caption{Study 2: AI and human decision-makers.}
         \label{fig:agent1}
     \end{subfigure}
     \caption{Differences in responsibility attribution to AI programs and humans for bail decisions. $^*$\(p<.05\), $^{**}$\(p<.01\), $^{***}$\(p<.001\).}
     \Description{Differences in responsibility attribution to AI programs and humans for bail decisions. Human agents were attributed more present-looking and forward-looking responsibility notions than AI agents. Backward-looking notions and responsibility-as-answerability were similarly ascribed to both human and AI agents.}
     \label{fig:agent}
\end{figure*}

Our primary goal was to examine how people attribute responsibility to human and AI agents in high-stakes scenarios. To quantify the difference, we used a multivariate linear mixed model that included a random-effects term to control for each participant. This allowed us to account for repeated measures, i.e., explicitly model that each participant responded to questions on eight distinct defendants. We use the standard .05 level of significance. In all models, we use our adapted scale of pre-attitude towards AI systems as a control variable.
Figure~\ref{fig:agent} shows the results. The annotated numbers indicate the differences and significance levels between the two agents. We report the full regression coefficients in Table~\ref{tab:regression_results} in the Appendix.

Both Study 1 and Study 2 show consistent differences in responsibility attribution between agents, regardless of whether they informed a human judge (Study 1) or decided by themselves (Study 2).
We note subtle differences in how people attribute responsibility to AI and humans. The first four responsibility concepts are correlated; the notions addressing tasks, supervisory roles, and the skills needed to assume them show a meaningful difference between agent types. The respondents attributed more of these notions of responsibility to humans than to AIs.

Responsibility-as-answerability exhibits a marginal difference with respect to the agent type that assisted human judges in bail decisions; however, the same trend was not observed in Study 2.
Nevertheless, our results suggest that humans and AI are judged similarly responsible with respect to causality, blame, and liability for bail decisions. Moreover, human decision-makers are praised to a considerably larger degree than AI decision-makers, although the same effect was not observed for human and AI advisors.

\section{Discussion}
\subsection{The Relation Between Notions of Responsibility}

So far, we have observed two clusters of responsibility concepts by their correlation.
The first cluster is composed of responsibility-as-task, authority, power, and obligation --- all of which were attributed to a greater degree to humans than AI systems
($\Delta$\textgreater0.206, $p$\textless.001). The first three are descriptive and focus on one's tasks (i.e., task, authority) and the necessary skills for their completion (i.e., power). Furthermore, responsibility-as-obligation is related to responsibility-as-task in prescribing a specific goal to the agent; it differs from the latter, however, in setting a supervisory role towards the task, rather than specifying that one should be the one to complete it.

The second cluster includes causal responsibility, blame, praise, and liability --- all of which were attributed to a similar degree to humans and AI.
This finding is in line with previous work on blame assignment, highlighting the significance of causality in people's ascription of blame and punishment. Human subject studies suggest that blame attribution is a two-step process; it is initiated by a causal connection between an agent's action and its consequences and is followed by evaluating its mental states, i.e., intentions~\cite{cushman2008crime}. Malle et al.~\cite{malle2014theory} have also proposed a theory of blame that is dependent on the causal connection between an agent and a norm-violating event. Our data similarly reveal such a relationship, even when controlling for the advice given, bail decision, or re-offense.  

Concerning the phrasing styles, our experiment design addressed responsibility-as-liability as the duty to compensate those harmed by a wrongful action. However, previous work on the connection between liability (i.e., punishment) and causality focuses on the retributive aspect of punishment~\cite{cushman2008crime}, often drawing a connection between punishment and blame. Therefore, we do not posit that people's ascription of liability is solely dependent on causality determinations. We hypothesize that the low assignment of liability is due to the current study's bail decision-making context. For instance, those wrongfully convicted do not receive any compensation for years spent in prison in at least 21 US states~\cite{cbscompensation}. Hence, people might not believe that compensation is needed or deserved, or attribute this notion of responsibility to other entities, such as the court or the government, leading to a lower ascription of liability to the advisor or decision-maker.

Our findings indicate that participants who were presented with responsibility statements addressing the violation or protection of a defendant's rights (e.g., ``It is the AI program’s task to protect the rights of the defendant'') were assigned higher responsibility levels across all notions. We posit that this effect results from the control that judges (humans and AIs) have over the consequences of their advice and decisions. Although a judge's decision can directly affect a defendant's rights depending on the appropriateness of one's jailing, preventing re-offenses is a complex task that encompasses diverse factors, such as policing and the defendant's decision to re-offend. 


\subsection{Humans Are More Responsible for Their Tasks Than AI Programs}

Participants perceived human judges and advisors as more responsible for their tasks than their AI counterparts (see the leftmost bars in Figure~\ref{fig:agent}). Humans {are} responsible for the tasks they are assigned, e.g., preventing re-offenses because they are in charge (i.e., authority) and have the skills necessary for completing them (i.e., power). These agents should either oversee (i.e., obligation) these tasks or take the lead (i.e., task). On the other hand, AI systems are ascribed lower levels of all these responsibility notions.

The meanings of responsibility addressing the attribution of tasks and their requirements are descriptive in the sense that they should be addressed in the present tense~\cite{davis2012ain}, e.g., one \emph{is} responsible for a task, or \emph{is} in charge of it. Although descriptive and present-looking, these notions lead to the prescription of forward-looking responsibilities, such as an obligation. For instance, to be responsible for a specific task because one has the authority and necessary skills prescribes that one should see to it that the task is completed, i.e., an obligation is prescribed, through consequentialist, deontological, or virtue-based routes~\cite{van2015moral}.

Participants attributed lower levels of authority and power to AI. This indicates that these systems are not thought to possess the necessary abilities to make decisions and advise such high-stakes decisions. Therefore, it {is not deemed} the AI program's responsibility to complete the assigned task or see it to be fulfilled.

\subsection{The Need for Explanations}

One of the prominent findings of this work is the need of interpretable AI systems. Although our participants assign a marginally lower level of responsibility-as-answerability for AI advisors vis-à-vis their human counterparts ($\Delta$=0.167, $p$\textless.05), they believe they should justify their decisions to the same extent as human judges, particularly if they are to make the final bail decision ($p$\textgreater.05).

Moreover, our results suggest that an AI without a human-in-the-loop, i.e., AI judges in Study 2, could be held at the same level of scrutiny as human decision-makers for their decisions. 
This finding may imply that deploying black box AI in high-stakes scenarios, such as bail decision-making, will not be perceived well by the public. There exists empirical evidence that people might be averse to machines making moral decisions~\cite{bigman2018people}. Previous work has not controlled for a system's interpretability, and therefore such trends might either \textit{i}) be caused by the lack of explanations or \textit{ii}) be aggravated if people become aware that AI systems cannot justify their moral decisions.

Judges should base their decisions on facts and be able to explain why they made such decisions. AI systems should also be capable of justifying their advice and decision-making process according to our results. This finding demonstrates the significance of these systems' interpretability. Scholars have discussed the risks posed by the opacity of existing AI algorithms. They argue that understanding how these systems come to their conclusions is necessary for both safe deployment and wide adoption~\cite{floridi2018ai4people}. Explainable AI (XAI)~\cite{gunning2017explainable} is a field of computer science that has been given much attention in the community~\cite{gilpin2018explaining}, and our results suggest that people agree with its importance.

Previous work has found that one's normative and epistemological values influence how explanations are comprehended~\cite{lombrozo2009explanation}. Explanations involve both an explainer and explainee, meaning that conflicts might arise concerning how they are evaluated~\cite{mittelstadt2019principles}. Therefore, we also posit that future work should delve deeper into what types of explanations the general public expects from AI systems. We highlight that those who are in charge of developing interpretable systems should not try to ``nudge'' recipients so they can be manipulated~\cite{lipton2018mythos}, e.g., for agency laundering~\cite{rubel2019agency}.

\subsection{AI and Human Agents Are Similarly Responsible for Consequences}


The four rightmost bars in Figure~\ref{fig:agent} suggest that AI and human agents are ascribed similar levels of backward-notions of responsibility, namely blame, liability, praise, and causal responsibility. 

\subsubsection{The Relation Between Causality and Blame}
A model that can explain our blameworthiness results is the Path Model of Blame, which proposes that blame is attributed through nested and sequential judgments of various aspects of the action and its agent~\cite{malle2014theory}. After identifying a norm-violating event, the model states that one judges whether the agent is causally connected to the harmful outcome. If this causal evaluation is not successful, the model assigns little or no blame to the agent. Otherwise, the blamer evaluates the agent's intentionality. If the action is deemed intentional, the blamer evaluates the reasons behind it and ascribes blame accordingly. For unintentional actions, however, one evaluates whether the agent should have prevented the norm-violating event (i.e., had an obligation to prevent it) and could have done so (i.e., had the skills necessary), hence blaming the agent depending on the evaluation of these notions.

Our results from both studies show that AI and human agents are blamed to a similar degree. These findings agree with the Path Model of Blame, which proposes causality as the initial step for blame mitigation. The model proposes that one can mitigate blame by i) challenging one's causal connection to the wrongful action or ii) defending that it does not meet moral eligibility standards. We posit that the first excuse can explain why people blame human and AI advisors and decision-makers similarly. As their causal connection to the consequence is deemed alike, they are attributed to similar blame levels. Challenging one's causal effect in an outcome has also been discussed as a possible excuse to avoid blame by other scholars~\cite{van2011relation}.

\subsubsection{Praise in AI-Assisted Bail Decisions}
The extent to which praise was assigned to human and AI agents varied depending on whether one was an advisor or a decision-maker. Even though Study 1 shows no difference between the two ($p$\textgreater.05), human decision-makers were more highly praised than AIs in Study 2 ($\Delta$=0.461, $p$\textless.001). Previous work has proposed praise as a positive reinforcement~\cite{kazdin1978history} and a method through which one might convey information about its values and expectations to the praisee~\cite{delin1994praise}. 

Regarding the difference between advisors and decision-makers, we posit that the differences between human agents are caused by the level of control the latter has over its decision outcomes. Although an advisor influences the final decision, the judge is the one who acts on it and, hence, deserves praise. Moreover, taking praise as positive reinforcement, praising the decision-maker over an advisor might have a bigger influence over future outcomes.

However, our results also indicate that AI decision-makers are not praised to the same level as human judges.
Taking praise as a method of conveying social expectations and values, we highlight that people might perceive existing praising practices as inappropriate for AI. Similarly to the arguments against holding AI responsible per se, focusing on the fact that they do not have mental states required for existing responsibility practices~\cite{sparrow2007killer,torrance2008ethics}, praising an AI might lose its meaning if done as if it were towards humans.

The same argument could also be applied to the practice of blame~\cite{danaher2016robots}. If the general public believes praising an AI system does not make sense, people might perceive blameworthiness similarly, contradicting our results. However, studies have shown a public impulse to blame, driven by the desire to express social values and expectations~\cite{carlsmith2002we}. Psychological evidence further suggests that humans are innate retributivists~\cite{carlsmith2008justifying}. Likewise, HCI research has found that people attribute blame to robotic agents upon harm, particularly if they are described to be autonomous and serve the main cause of harm~\cite{malle2015moral,furlough2019attributing,kim2006should}.
Hence, there is no contradiction in people attributing blame to AI systems for harms, although they should not be praised for opposing consequences.

\subsubsection{Liability as Compensation}
Our findings indicating that AI and human agents should be held liable to a similar level goes against previous work, which has found that people attribute punishment to AI systems to a lesser degree than their human counterparts~\cite{limachi2020}. Punishment fulfills many societal goals, such as making victims whole, the satisfaction of retributive feelings, and offenders' reform. In the current study, we address one of these functions and phrase liability as the responsibility to compensate those harmed (i.e., make victims whole). Therefore, our results do not directly contradict earlier findings that had addressed punishment in its wide definition.

The results from our initial exploratory analysis in Section~\ref{sec:exploratory} show that trends found between causality and blame attributions across different phrasing styles do not directly transfer to liability judgments. Hence, we do not posit that similar causality judgments can explain the similar attribution of liability to AI and humans as in the case of blame. Still, we instead hypothesize that it results from two different factors based on our phrasing styles. 

Regarding the statements addressing the prevention of re-offenses, we posit that the lower attribution of liability to both agents is caused by a variation of the ``problem of many hands.'' ~\cite{van2015moral} Preventing defendants from re-offending does not rely solely on a judge's decision but encompasses many other factors as discussed above. Therefore, liability is distributed across various entities, such as the government and the court per se. Addressing the statements focusing on protecting defendants' rights, we hypothesize that people do not expect defendants to be compensated if their rights are violated. As examined above, much of the US legislature does not compensate those who have been unjustly incarcerated~\cite{cbscompensation}. The respondents did not believe those harmed should, or even could, be made whole for the violation of their rights, and hence, both AI and human agents are attributed low and similar levels of liability.

\section{Implications}

Our findings indicate that people believe humans {are}, and {should be}, responsible for the assigned tasks, regardless of whether they are advisors or decision-makers. Our respondents perceive humans as having the skills necessary to complete these tasks, being in charge of them, and being able to ensure that they are completed. The responsibility notions that were attributed to human agents to a greater extent than to AIs are present- and forward-looking in the sense that they are descriptive, i.e., by stating a fact, and prescribe obligations. It is important to note that users of AI systems are also responsible in a backward-looking fashion such that they should also be held responsible for the outcomes of their advice and decisions. Therefore, our findings agree with scholars who propose that users (and designers) should take responsibility for their automated systems' actions and consequences~\cite{champagne2015bridging,nyholm2018attributing}.

Nonetheless, our study shows that AIs could also be held responsible for their actions. Taking morality as a human-made construct~\cite{theodorouartificial}, it may be inevitable to hold AI systems responsible alongside their users and designers so that this formulation is kept intact. Viewing responsibility concepts as social constructs that aim to achieve specific social goals, attributing backward-looking notions of responsibility to AI systems might emphasize these goals~\cite{stahl2006responsible}. Our study indicates these practices might not need to focus on compensating those harmed by these systems given the low attribution of liability to all agents.\footnote{This finding does not imply that those harmed should not be compensated, but rather that respondents do not attribute this responsibility to AI systems per se. Some scholars propose that other stakeholders should take this responsibility~\cite{vcerka2015liability}, mainly because automated agents are not capable of doing so~\cite{bryson2017and}.} We instead hypothesize that people might desire to hold these entities responsible for retributive motives, such as satisfying their needs for revenge~\cite{mulligan2017revenge} and bridging the retribution gap~\cite{danaher2016robots}, as a result of human nature~\cite{cushman2008crime}. It is important to note that AI systems might not be appropriate subjects of (retributive) blame~\cite{sparrow2007killer,danaher2016robots}, i.e., scholars argue that blaming automated agents would be wrong and unsuccessful. Future research can address which functions of responsibility attribution would satisfy this public attribution of backward-looking responsibilities to AI systems. Future studies can also address scenarios in which blame could be attributed to a higher degree, e.g., those with life-or-death consequences, such as self-driving vehicles and AI medical advisors.

A common concern raised by scholarly work is that blaming or punishing an AI system might lead to social disruptions. From a legal perspective, attributing responsibility to these systems might obfuscate designers and users' roles, creating human liability shields~\cite{bryson2017and}, i.e., stakeholders might use these automated systems as a form of protecting themselves from deserved punishment. Another possible issue is ``agency laundering,'' in which the systems' designer distances itself from morally suspect actions, regardless of intentionality, by blaming the algorithm, machine, or system~\cite{rubel2019agency}. This form of blame-shifting has been observed, for example, when Facebook called out its algorithm for autonomously creating anti-semitic categories in its advertisement platform~\cite{propublica_fb,tsamados2020ethics}. We highlight that any responsibility practice towards AI systems should not blur the responsibility prescribed and deserved by their designers and users. Our findings suggest that autonomous algorithms alone should not be held responsible by themselves, but rather \emph{alongside} other stakeholders, so these concerns are not realized.

\section{Concluding Remarks}

This paper discussed the responsibility gap posed by the deployment of autonomous AI systems~\cite{matthias2004responsibility} and conducted a survey study to understand how differently people attribute responsibility to AI and humans. As a case study, we adapted vignettes from real-life algorithm-assisted bail decisions and employed a within-subjects experimental design to obtain public perceptions on various notions of moral responsibility. We conducted two studies; the former illustrated a realistic scenario in which AI advises human judges, and the latter described a fictional circumstance where AI is the decision-maker itself.

The current study focused on AI systems currently being used to advise bailing decisions, which is an important yet specific application of these algorithms. Therefore, our results might not be generalizable to all possible environments. For instance, some of our results partly conflict with previous work addressing self-driving vehicles~\cite{awad2020drivers} and medical systems~\cite{limachi2020}. Studies such as ours should be expanded to diverse AI applications, where they are used both in-the-loop (as in Study 1) and autonomously (as in Study 2). People have different opinions regarding how (and where) these systems should be deployed in relation to how autonomous they should be~\cite{lubars2019ask}, which should affect how they ascribe responsibility for their actions.

Study 1 was designed so that the judge's decision always followed the advice given to reduce complexity in the vignette design. However, future studies on similar topics should also consider scenarios in which AI systems and their human supervisors disagree. For instance, if a human judge chooses to disagree with advice, some of the advisor's responsibilities might be shifted towards the decision-maker regardless of the advisor's nature. In our case study, human-AI collaborations are such that there exists an AI-in-the-loop; future work should address other collaboration variations, such as human-in-the-loop, i.e., humans assisting machines.

The current research considered eight notions of responsibility from related work. We recognize that other meanings of responsibility could be further considered, such as virtue-based notions where one might call an entity responsible in that it prescribes an evaluation of one's traits and dispositions~\cite{shoemaker2011attributability,tigard2020responsible}. These notions have been widely agreed upon as incompatible with AI systems due to their lack of metaphysical attributes~\cite{sparrow2007killer,champagne2015bridging,torrance2008ethics}. Nevertheless, our research has found key clusters of responsibility notions concerning AI and human agents, opening further research directions.

Our exploratory analysis identified two clusters of responsibility notions. One cluster encompasses meanings related to the attribution of tasks and obligations (i.e., responsibility-as-task, as-obligation), their necessary skills (i.e., responsibility-as-power), and the ascription of authority (i.e., responsibility-as-authority). The other cluster includes meanings related to causal determinations (i.e., responsibility-as-cause) and backward-looking responsibility notions (i.e., blame, praise, and liability).  

As our results demonstrate, people may hold AI to a similar level of moral scrutiny as humans for their actions and harms. Our respondents indicate that they expect decision-makers and advisors to justify their bailing decisions regardless of their nature. Our findings highlight the importance of interpretable and explainable algorithms, particularly in high-stakes scenarios, such as our case-study. Finally, this study also showed that people judge AI and humans differently with respect to certain notions of responsibility, particularly those addressing present- and forward-looking meanings, such as responsibility-as-task and as-obligation. However, we have also found that people attribute similar levels of causal responsibility, blame, and liability to AI and human advisors and decision-makers for bail decisions.


\bibliographystyle{ACM-Reference-Format}
\bibliography{sample-base}


\begin{acks}
    This work was supported by the Institute for Basic Science (IBS-R029-C2).
\end{acks}

\begin{appendix}
\newpage
\section{Appendix}

\begin{figure*}[htb!]
    \centering
    \begin{subfigure}[]{.49\textwidth}
        \includegraphics[width=\textwidth]{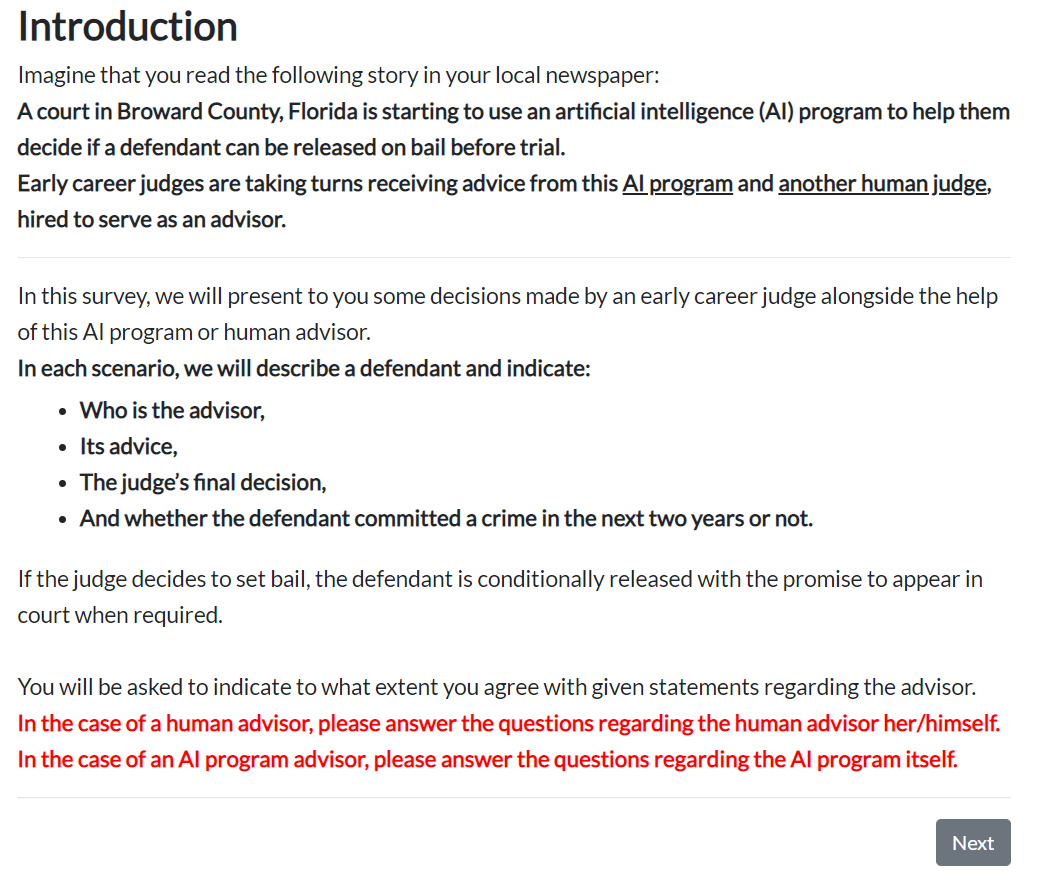}
        \caption{Study introduction presenting the scenario where AI systems are being used for bail decisions.}
    \end{subfigure}
    \begin{subfigure}[]{.49\textwidth}
         \includegraphics[width=\textwidth]{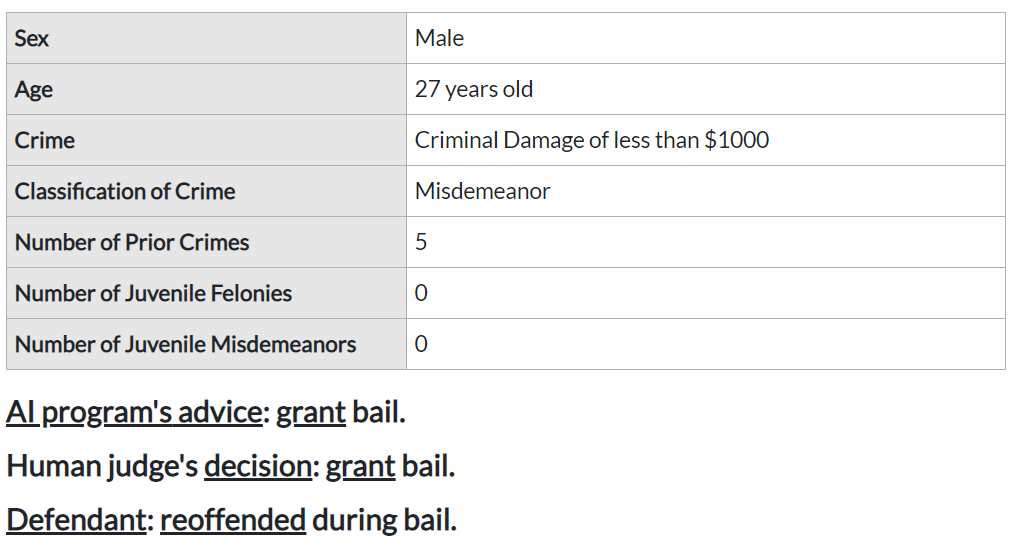}
         \caption{Vignette presented to survey participants introducing a defendant, whether they have re-offended, and the stakeholders' decisions and advices.}
    \end{subfigure}
    \begin{subfigure}[]{\textwidth}
         \centering
         \vspace{10px}
         \includegraphics[width=.70\textwidth]{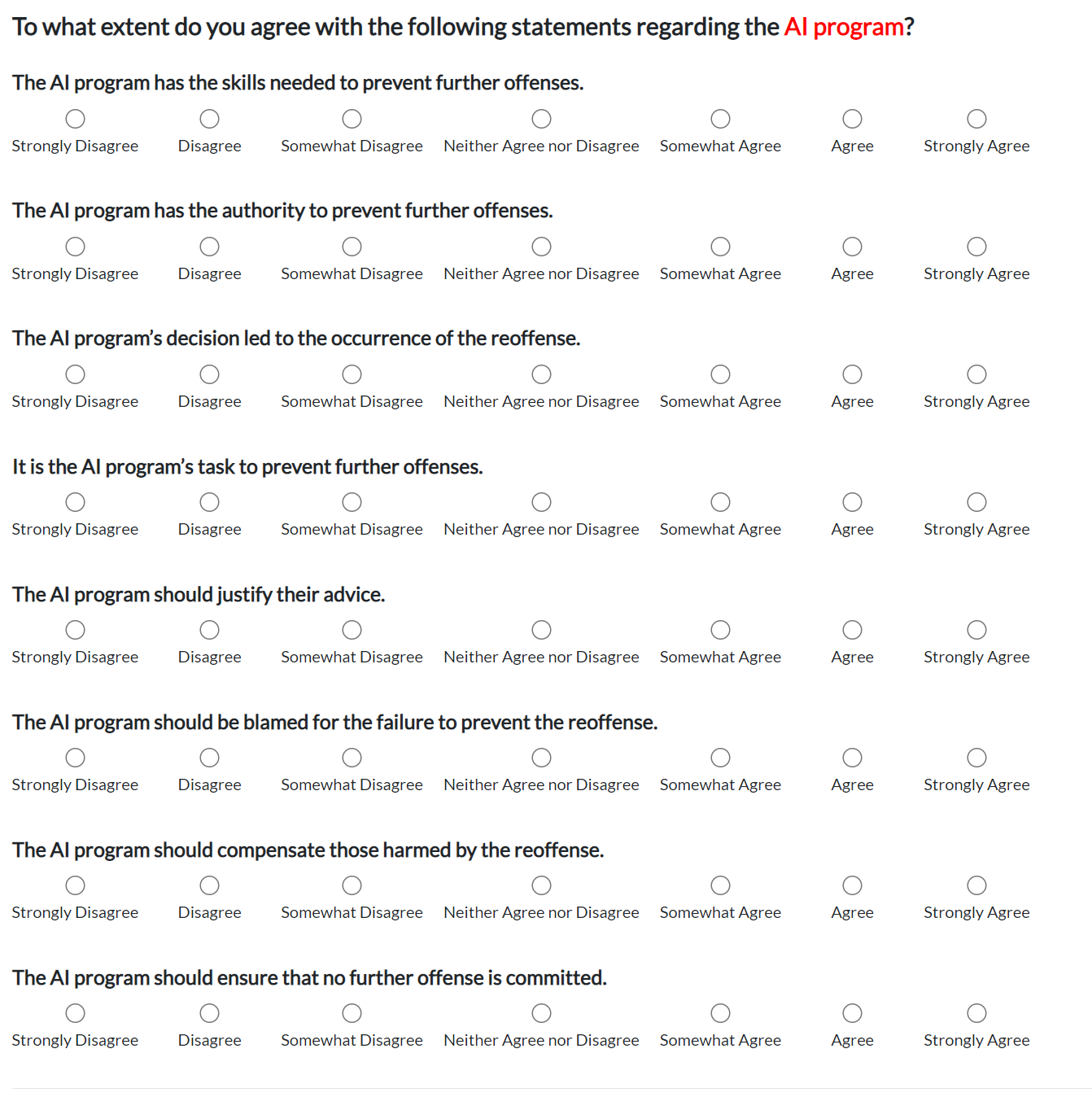}
         \caption{Attribution of the eight notions of moral responsibility to the advisor in Study 1 (or decision-maker in Study 2).}
    \end{subfigure}
    \caption{Example screenshots of the survey instrument used for Study 1. The study is available at \url{https://thegcamilo.github.io/responsibility-compas/}.}
    \label{fig:survey_instrument}
\end{figure*}

\begin{figure*}[t]
     \begin{subfigure}[b]{.8\textwidth}
         \centering
         \includegraphics[width=0.75\textwidth]{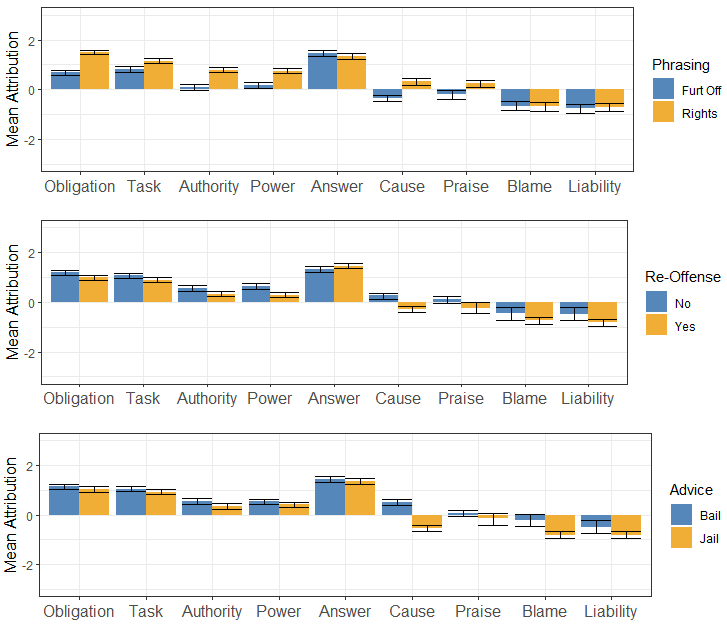}
         \caption{Study 1: AI and human advisors.}
         \label{fig:agent2}
     \end{subfigure}
    \begin{subfigure}[b]{.8\textwidth}
         \centering
         \includegraphics[width=0.75\textwidth]{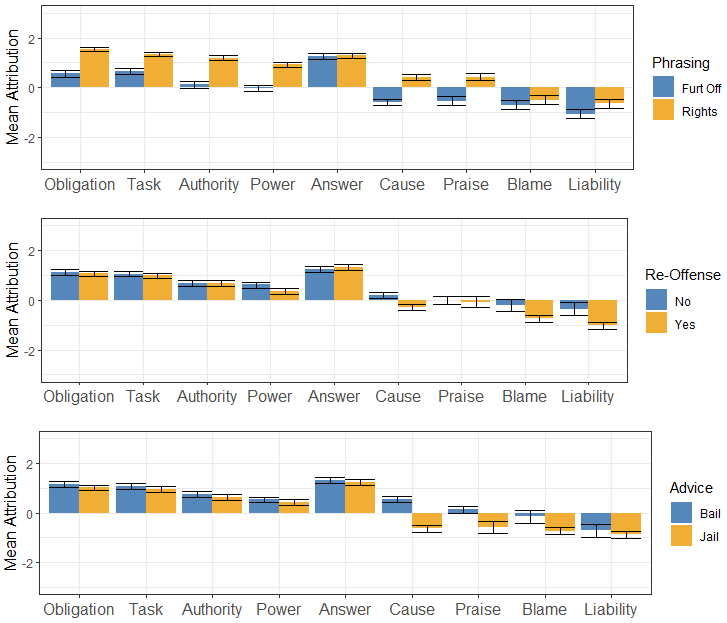}
         \caption{Study 2: AI and human decision-makers.}
         \label{fig:agent1}
     \end{subfigure}
     \caption{Attribution of responsibility for bail decisions depending on how the statements were phrased, recidivism, and advice/decision.}
     \label{fig:app}
\end{figure*}

\begin{table*}[htb!]
\centering
\small
\begin{tabulary}{\linewidth}{l|l|l}
\toprule
     \textbf{Notion} & \textbf{Phrasing} & \textbf{Statements} \\
     \hline
     \multirow{2}{*}{Responsibility-as-task} & Further Offense & It is the (agent)’s task to prevent further offenses. \\
     & Rights & It is the (agent)’s task to protect the rights of the defendant. \\ 
     \hline
     \multirow{2}{*}{Responsibility-as-authority} & Further Offense & The (agent) has the authority to prevent further offenses. \\
     & Rights & The (agent) has the authority to protect the rights of the defendant. \\ 
     \hline
     \multirow{2}{*}{Responsibility-as-power} & Further Offense & The (agent) has the skills needed to prevent further offenses. \\
     & Rights & The (agent) has the skills needed to protect the rights of the defendant. \\ 
     \hline
     \multirow{2}{*}{Responsibility-as-obligation} & Further Offense & The (agent) should ensure that no further offense is committed. \\
     & Rights & The (agent) should ensure that the rights of the defendant are protected. \\
     \hline
     \multirow{2}{*}{Responsibility-as-answerability} & Further Offense & The (agent) should justify their advice/decision. \\
     & Rights & The (agent) should justify their advice/decision. \\ 
     \hline
    \multirow{2}{*}{Responsibility-as-cause} & Further Offense & The (agent)'s decision led to the occurrence/prevention of the reoffense. \\
     & Rights & The (agent)’s decision led to the violation/protection of the rights of the defendant. \\ 
     \hline
    \multirow{2}{*}{Responsibility-as-blame/praise} & Further Offense & \vtop{\hbox{\strut The (agent) should be blamed/praised for the failure to prevent/prevention}\hbox{\strut of the reoffense.}} \\
     & Rights & \vtop{\hbox{\strut The (agent) should be blamed/praised for the violation/protection of the rights}\hbox{\strut of the defendant.}} \\ 
     \hline
     \multirow{2}{*}{Responsibility-as-liability} & Further Offense & The (agent) should compensate those harmed by the reoffense. \\
     & Rights & The (agent) should compensate the defendant for violating their rights. \\
\bottomrule
\end{tabulary}
\caption{Statements addressing all responsibility notions presented to participants in Study 1 and Study 2. (Agent) is either ``AI program,'' ``human advisor,'' or ``human judge'' depending on the agent and the study. The statements addressing responsibility-as-liability were shown if \textit{i}) the defendant re-offended and the phrasing style addressed the prevention of re-offenses, or \textit{ii}) the defendants were denied bail and did not re-offend within two years while the statements focused on the protection of their rights. The phrases tackling praise and blame were presented depending on the advice/decision and recidivism. The phrasing column indicates how statements were phrased depending on which function of the bail decision they stressed: preventing further offenses (Further Offense) or protecting the defendant's rights (Rights).}
\label{tab:statements}
\end{table*}

\def\sym#1{\ifmmode^{#1}\else\(^{#1}\)\fi}
\begin{table*}
\footnotesize{
\begin{tabular}{l*{2}{c}}
\toprule
            &\multicolumn{1}{c}{(1)}&\multicolumn{1}{c}{(2)}\\
            &\multicolumn{1}{c}{Decision-Maker}&\multicolumn{1}{c}{Advisor}\\
\hline
\textbf{answer}&                     &                     \\
agent\_human &     -0.0232         &       0.168\sym{*}  \\
advice\_jail &     -0.0722         &     -0.0800         \\
defendant\_reoffended&      0.0696         &       0.130         \\
phrasing\_rights&      0.0308         &      -0.102         \\
control     &      0.0103         &      -0.119         \\
intercept      &       1.254\sym{***}&       1.501\sym{***}\\
\hline
\textbf{authority}&                     &                     \\
agent\_human &       0.885\sym{***}&       0.753\sym{***}\\
advice\_jail &      -0.122\sym{*}  &      -0.193\sym{**} \\
defendant\_reoffended&    -0.00129         &      -0.233\sym{***}\\
phrasing\_rights&       1.052\sym{***}&       0.702\sym{***}\\
control     &      -0.150         &     -0.0790         \\
intercept      &     -0.0410         &      0.0360         \\
\hline
\textbf{blame} &                     &                     \\
agent\_human &      0.0258         &       0.107         \\
advice\_jail &      -1.128\sym{***}&      -0.878\sym{***}\\
defendant\_reoffended&      -0.572\sym{***}&      -0.456\sym{**} \\
phrasing\_rights&       0.500\sym{*}  &       0.190         \\
control     &       0.239\sym{*}  &     -0.0312         \\
intercept      &      0.0484         &       0.226         \\
\hline
\textbf{cause} &                     &                     \\
agent\_human &      0.0863         &       0.115         \\
advice\_jail &      -1.179\sym{***}&      -1.018\sym{***}\\
defendant\_reoffended&      -0.460\sym{***}&      -0.550\sym{***}\\
phrasing\_rights&       1.020\sym{***}&       0.665\sym{***}\\
control     &      0.0278         &     -0.0686         \\
intercept      &       0.136         &       0.468\sym{**} \\
\hline
\textbf{liability} &                     &                     \\
agent\_human &     -0.0541         &       0.122         \\
advice\_jail &      -0.722\sym{***}&      -0.541\sym{**} \\
defendant\_reoffended&      -0.553\sym{***}&      -0.466\sym{**} \\
phrasing\_rights&       0.530\sym{*}  &      0.0999         \\
control     &       0.232\sym{*}  &      0.0355         \\
intercept      &      -0.481         &      -0.145         \\
\hline
\textbf{obligation} &                     &                     \\
agent\_human &       0.206\sym{***}&       0.279\sym{***}\\
advice\_jail &      -0.144\sym{**} &      -0.124\sym{*}  \\
defendant\_reoffended&     -0.0438         &      -0.216\sym{***}\\
phrasing\_rights&       0.987\sym{***}&       0.815\sym{***}\\
control     &     -0.0614         &     -0.0786         \\
intercept      &       0.642\sym{***}&       0.814\sym{***}\\
\hline
\textbf{power} &                     &                     \\
agent\_human &       0.799\sym{***}&       0.649\sym{***}\\
advice\_jail &      -0.101         &      -0.126         \\
defendant\_reoffended&      -0.271\sym{***}&      -0.366\sym{***}\\
phrasing\_rights&       0.939\sym{***}&       0.572\sym{***}\\
control     &     -0.0759         &      -0.118         \\
intercept      &      -0.119         &       0.263         \\
\hline
\textbf{praise} &                     &                     \\
agent\_human &       0.461\sym{***}&       0.128         \\
advice\_jail &     -0.0611         &      0.0969         \\
defendant\_reoffended&      -0.990\sym{***}&      -0.941\sym{***}\\
phrasing\_rights&       1.435\sym{***}&       0.950\sym{***}\\
control     &      0.0277         &      -0.120         \\
intercept      &      -0.784\sym{***}&      -0.148         \\
\hline
\textbf{task} &                     &                     \\
agent\_human &       0.282\sym{***}&       0.334\sym{***}\\
advice\_jail &      -0.117\sym{*}  &      -0.111         \\
defendant\_reoffended&     -0.0683         &      -0.164\sym{**} \\
phrasing\_rights&       0.686\sym{***}&       0.345\sym{**} \\
control     &     -0.0417         &      0.0128         \\
intercept      &       0.667\sym{***}&       0.760\sym{***}\\
\bottomrule
\end{tabular}
}
\caption{Coefficients from the multivariate mixed effects model presented in Section~\ref{sec:regression}. $^*$\(p<.05\), $^{**}$\(p<.01\), $^{***}$\(p<.001\).}
\label{tab:regression_results}
\end{table*}



\end{appendix}

\end{document}